\documentclass[conference]{IEEEtran}
\IEEEoverridecommandlockouts

\usepackage{cite}
\usepackage{amsmath,amssymb,amsfonts}
\usepackage{algorithm}
\usepackage{algorithmic}
\usepackage{graphicx}
\usepackage{textcomp}
\usepackage{xcolor}
\usepackage{hyperref}
\usepackage{listings}

\lstdefinestyle{sqlstyle}{
    language=SQL,
    basicstyle=\ttfamily\footnotesize,
    keywordstyle=\color{blue}\bfseries,
    stringstyle=\color{red},
    commentstyle=\color{gray}\itshape,
    morecomment=[l]{--},
    breaklines=true,
    showstringspaces=false,
    columns=fullflexible
}

\def\BibTeX{{\rm B\kern-.05em{\sc i\kern-.025em b}%
\kern-.08em T\kern-.1667em\lower.7ex\hbox{E}\kern-.125emX}}
\usepackage{listings}
\usepackage{courier} 
\begin{document}

\title{Precision over Noise: Tailoring S3 Public Access Detection to Reduce False Positives in Cloud Security Platforms}

\author{
  \IEEEauthorblockN{Dikshant\textsuperscript{a}}
  \IEEEauthorblockA{\textsuperscript{a}7-Eleven, Bengaluru, Karnataka, India\\
  Email: 27dikshant@gmail.com}
  \and
  \IEEEauthorblockN{Geetika Verma\textsuperscript{b}}
  \IEEEauthorblockA{\textsuperscript{b}RMIT University, Melbourne, Australia\\
  Email: geetika.verma@rmit.edu.au}
}

\maketitle

\maketitle

\begin{abstract}
Excessive and spurious alert generation by cloud security solutions is a root cause of analyst fatigue and operational inefficiencies. In this study, the long-standing issue of false positives from publicly accessible alerts in Amazon S3, as generated by a licensed cloud-native security solution, is examined. In a simulated production test environment, which consisted of over 1,000 Amazon S3 buckets with diverse access configurations, it was discovered that over 80\% of the alerts generated by default rules were classified as false positives, thus demonstrating the severity of the detection issue. This severely impacted detection accuracy and generated a heavier workload for analysts due to redundant manual triage efforts. For addressing this problem, custom detection logic was created as an exercise of the native rule customization capabilities of the solution. A unified titled ``S3 Public Access Validation and Data Exposure'' was created in an effort to consolidate different forms of alerts into one, context-aware logic that systematically scans ACL configurations, bucket policies, indicators of public exposure, and the presence of sensitive data, and then marks only those S3 buckets that indeed denote security risk and are publicly exposed on the internet with no authentication. The results demonstrate a significant reduction in false positives, more precise alert fidelity, and significant time saving for security analysts, thus demonstrating an actionable and reproducible solution to enhance the accuracy of security alerting in compliance-focused cloud environments.

\end{abstract}
\vspace{1em}

\section{Introduction}
The rapid adoption of cloud technology has led to a rise in security issues related to misconfiguration, especially in object storage offerings like Amazon Simple Storage Service (S3). Despite S3 providing an integrated suite of access controls such as bucket policies, access control lists (ACLs), and block public access settings, misconfigurations continue to present significant and high-impact risks.

Public exposure of S3 buckets remains a prime cause of inadvertent disclosure of confidential data such as access keys, API credentials, and proprietary information, leading to compliance failures and reputational damage.

Public evidence shows that misconfigured S3 permissions caused more than 2,600 publicly reported data breaches between 2018 and 2024. High-profile incidents include the 2021 Booz Allen Hamilton breach, where S3 buckets were found leaking sensitive documents and credentials \cite{booz2021}, the 2017 Accenture cloud misconfiguration \cite{accenture2017}, and the U.S. Army Intelligence \& Security Command exposure in 2017 \cite{usarmy2017}. These cases highlight the urgent need for reliable and efficient detection of S3 misconfigurations. The false alerts consisted of overly general Access Control List (ACL) checks, erroneous inferences of public exposure, and redundant policy violations. The impact of such over alerting is analyst burnout, increased triage requirements, and increased incident response time \cite{veriti2025}, \cite{musa2021}, \cite{saleh2022}, \cite{islam2022}.

While current academic research has examined methods such as anomaly detection using graph neural networks \cite{marbel2024}, contextual alert enrichment \cite{islam2022}, and noise filtering using machine learning methods \cite{lee2020}, the real world applicability of these methods in actual CNAPP production environments is limited. Current research emphasizes that the alert generation, relevance to business context gap remains a major barrier to effective cloud security posture management \cite{saqib2025}, \cite{oliver2024}, \cite{lucid2024}.

In order to address this challenge, this research presents a custom context aware detection rule that aggregates over 20 unique and often ambiguous forms of alerts into a unified, high-fidelity alert type called ``S3 Public Access Validation and Data Exposure.'' This rule, when applied to a commercial CNAPP solution, unifies various aspects of access control, such as ACLs, policy and validation, to identify legitimate public exposure risks.

We validated the precision and performance gain of this technique through controlled testing, with significant reductions in false positives and manual triage effort. Our outcomes indicate the requirement for customized detection techniques and present a reproducible method for optimizing alerts in large, compliance sensitive AWS environments \cite{veriti2025}--\cite{fotiou2022}, \cite{levelblue2024}, \cite{lucid2024},\cite{cyrebro2024}.

This research addresses this gap by introducing a custom, context-aware detection rule that merges over 20 ambiguous alert types into a unified, high-fidelity rule titled ``S3 Public Access Validation and Data Exposure.`` The proposed rule improves alert precision, embeds business logic natively into CNAPP queries, and has been empirically validated in a production-scale environment.

\section{Related Work}

As cloud-native businesses expand, the problem of dealing with false positives on cloud security platforms has become increasingly significant. Numerous studies have examined the structural limitations of Cloud Security Posture Management (CSPM) and Cloud-Native Application Protection Platforms (CNAPPs), especially in the context of Amazon S3 public access alerts \cite{sharma2020, sun2021, fotiou2022}.

\subsection{Alert Fatigue and Over-Detection in CNAPP Tools}

Alert fatigue: when security analysts stop paying attention to alerts because there are too many false positives and most aren’t useful is a well-known problem. Nations \cite{nations2025} and Veriti Research \cite{veriti2025} emphasized that the high volume of false positives generated by cloud security products results in reduced responsiveness and misallocation of security team resources. A Dazz.io recent remediation trend report \cite{dazz2024} found that over 45\% of an analyst's time is consumed triaging low-quality alerts, with response to actual threats being delayed.

Likewise, research by CYREBRO~\cite{cyrebro2024} and LevelBlue~\cite{levelblue2024} points out legacy, rule-based detection capabilities cannot match the complexity of modern cloud infrastructures like multi-account deployments, dynamic policies, and identity-based access controls. Our own testing in a licensed CNAPP validated this trend: default detection heuristics for Amazon S3 public access misconfigurations generated over 80\% false positives on enterprise scale workloads. They validate concerns expressed in recent academic and industry research that default heuristics employed by most CNAPPs are not fine-grained enough to prioritize actual risks~\cite{sharma2020,sun2021}.

\subsection{False Positive Reduction Strategies}

Research has proposed various strategies to reduce alert noise. Saleh and Gia \cite{saleh2022} proposed contextual validation using environmental metadata. Lee et al. \cite{lee2020} developed a classification-based machine learning model that filters alerts based on behavior. While these approaches improve precision, they often require complex pipelines or annotated data, limiting deployment feasibility in vendor-bound CNAPP environments.

Techniques like Carbon Filter \cite{oliver2024} and LUCID \cite{lucid2024} use clustering and ensemble filtering, respectively, but face integration and portability challenges.

\subsection{Enriching Alerts with Business Context}

Islam et al. \cite{islam2022} argued for alert enrichment using layers of business policy, resource sensitivity, and behavioral expectations. Our solution is well in the spirit of this thought, however, it differs from it in that it incorporates contextual logic within the detection layer using the CNAPP's native query language, bringing contextual logic to bear at the time of detection instead of as a subsequent processing step.

Wu and Shakir \cite{wu2021} explored anomaly detection methods, but these often sacrifice interpretability or operational simplicity. Our approach avoids the trade-off by designing high-specificity rules with care that business-aligned precision is feasible without extraneous telemetry or anomaly detection layers.

\subsection{Gaps in Existing Literature}

Despite growing recognition of false positives, literature still focuses largely on generic misconfiguration detection or ML proof-of-concepts. Our work addresses this gap by introducing a vendor-native, rule-level solution that enhances alert fidelity without third-party complexity or AI dependencies \cite{saqib2025, algomox2025}.

\begin{table*}[htbp]
\caption{Comparative Analysis of Prior Approaches and the Proposed Method}
\label{table:comparison-methods}
\centering
\begin{tabular}{|p{2.8cm}|p{3.4cm}|p{3.5cm}|c|c|c|c|}
\hline
\textbf{Approach} & \textbf{Technique} & \textbf{Tool/Framework} & \textbf{\shortstack{Business\\Context\\Awareness}} & \textbf{\shortstack{Ease of\\Implementation}} & \textbf{\shortstack{Platform\\Specific}} & \textbf{\shortstack{False Positive\\Reduction}} \\
\hline
Saleh \& Gia \cite{saleh2022} & Contextual validation via metadata & Custom contextual engine & Partial & Moderate & No & High \\
Lee et al. \cite{lee2020} & ML-based classification & Proprietary models & Low & Complex & No & Moderate \\
Islam et al. \cite{islam2022} & Contextual enrichment + policy mapping & Custom enrichment pipelines & High & Complex & No & High \\
Carbon Filter \cite{oliver2024} & Clustering and similarity scoring & Graph-based clustering & Low & High & No & Moderate \\
LUCID \cite{lucid2024} & Unified scan result filtering & Ensemble filtering layer & Low & High & No & Moderate \\
\textbf{Our Method (This Work)} & Custom rule logic using native CNAPP query & Licensed CNAPP + Custom Query Engine & High & Low & Yes & \textbf{Very High} \\
\hline
\end{tabular}
\end{table*}

Table~\ref{table:comparison-methods} presents a comparison of academic and industry approaches to reducing false positives in cloud security alerting, contrasted against our proposed methodology. While previous techniques, such as metadata-driven contextual verification \cite{saleh2022}, ML classifiers \cite{lee2020}, and clustering-based triage \cite{oliver2024} show potential, they often require substantial integration effort, third-party engines, or lack deployment-specific awareness.

By contrast, our approach is implemented entirely within the native query model of a licensed CNAPP. It takes advantage of business context such as access intent, sensitivity markers, and ACL settings at a rule level, thus greatly enhancing accuracy with the overhead of machine learning pipelines or custom enrichment infrastructure. This makes it extremely well-suited for production deployment on enterprise-scale AWS estates with high signal fidelity and low operational friction.

\begin{figure}[htbp]
    \centering
    \includegraphics[width=0.6\linewidth]{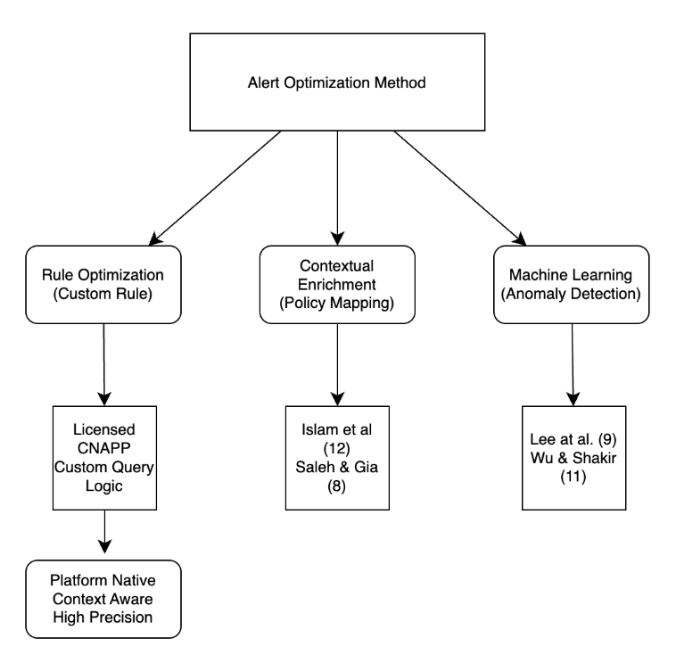} 
    \caption{Taxonomy of Alert Optimization Techniques}
    \label{fig:taxonomy}
\end{figure}

Figure~\ref{fig:taxonomy} illustrates a taxonomy of alert optimization methods for cloud security, categorized into three distinct groups: Rule Optimization, Contextual Enrichment, and Machine Learning-Based Detection. These categories are intended to characterize increasingly advanced methods that seek to reduce false positives in automated threat detection in the cloud.

Our proposed approach falls under the category of \textit{Rule Optimization}, which attempts to gain improved accuracy via native policy tuning rather than add-ons. Specifically, we leverage the native querying capability of a licensed CNAPP to insert business logic and contextual validation into the detection process natively. This is differentiated from \textit{Contextual Enrichment} methods relying on extrinsic metadata inputs or correlation processes, as well as from \textit{Machine Learning-Based Detection}, which typically requires considerable labeled data and infra for model training and inference.

By incorporating business awareness into detection logic itself, our solution is highly accurate and has low operational loads. It is most worth it for real-world cloud environments that must have quick, auditable, and interpretable security results irrespective of black-box or resource intensive systems.

To address these limitations, we introduce a native CNAPP rule that merges over 20 default alerts into a single detection logic that is contextually appropriate. The following section outlines our research questions and approach.

\section{Research Questions and Methodological Design}

In large-scale enterprise environments, Cloud-Native Application Protection Platforms (CNAPPs) are commonly used to detect misconfigurations and enforce security policies on services like Amazon S3. However, these platforms often rely on overly broad detection mechanisms, producing excessive alerts that overwhelm security teams.

While past research emphasizes external enrichment pipelines \cite{islam2022}, classifiers \cite{lee2020}, or AI-driven correlation tools \cite{oliver2024}, this study shows that improvements in alert precision can be achieved natively using the CNAPP's built-in query language. This approach simplifies adoption, ensures regulatory compatibility, and avoids the complexity of external integrations.

\subsection{Research Questions}

This work is driven by the following questions:

\begin{itemize}
    \item \textbf{RQ1:} Can a single, context-aware detection rule effectively replace more than 20 fragmented default alerts related to S3 public access, while maintaining or improving detection quality?
    \item \textbf{RQ2:} What measurable impact does the custom rule have on key metrics such as precision, alert reduction rate, and analyst triage workload?
    \item \textbf{RQ3:} To what extent can real-world business context be embedded directly into CNAPP-native detection logic without reliance on external enrichment or AI models?
    \item \textbf{RQ4:} Is the proposed methodology scalable and generalizable across complex, compliance-driven AWS environments, especially those handling sensitive workloads?
\end{itemize}

\subsection{Methodology Overview}

To evaluate the proposed rule’s effectiveness, we conducted a multi-phase experimental study within a production-grade AWS simulation environment consisting of over 1,000 S3 buckets. The study compares default CNAPP rules versus a unified, context-aware detection rule, implemented natively.

\subsubsection{Initial Assessment Using Standard S3 Detection Protocols}

Baseline with the default CNAPP ruleset was created, which included more than 20 S3 misconfiguration detection rules. The rules were created in an effort to identify problems like too-permissive ACLs, wildcard principals, and absence of Block Public Access (BPA) flags.

The baseline run contained a staggering number of alarms. Manual checking identified that the overwhelming majority were false alarms, such as misclassifications of buckets with internal service roles, valid IAM permissions, or compensating controls such as IP blocklists and conditional keys.

A common instance was a bucket with legacy ACLs that were functionally non-influential but nevertheless caused many overlapping rules to fire. Another instance was buckets designated with disabled BPA flags, although policies that enforced internal-only access.

\subsubsection{Unified Detection Logic: Consolidating Contextual Understanding}

To address the high rate of false positives, we built an in-house detection rule named \textit{S3 Public Access Validation and Data Exposure}. The rule integrates the logic of over 20 pre-built alerts and performs contextual analysis of S3 bucket security based on ACLs, bucket policies, BPA indicators, and metadata tags such as \texttt{SensitiveData}. In contrast to legacy rules, it alerts on global ACL grants (e.g., \texttt{AllUsers}, \texttt{AuthenticatedUsers}) and wildcard principals only when they don't have restrictive conditions. BPA indicators are employed as risk multipliers, not as single indicators. Buckets specifically labeled as sensitive override when risk indicators are present. Importantly, the rule fires only when multiple risk indicators occur concurrently, minimizing alert noise. It was built using the CNAPP native query language and tuned using iterative testing to attain accuracy and performance.

\subsubsection{Implementation Using Vendor-Native Query Language}

\subsection{Implementation Using CNAPP Native Query Language}

To convert the conceptual rule design model into an operational service, the vendor language was used to encode the combined detection logic. Its deployment required the creation of a composite query that would evaluate ACL entries, bucket policy statements, block public access (BPA) indicators, and indicators of exposure to business-sensitive information together. The language's syntax allowed compound logical operators, i.e., \texttt{AND}, \texttt{OR}, and \texttt{NOT}, to be used against multiple hierarchical resource attributes, thereby allowing the creation of a high-fidelity and accurate detection rule.

In query logic, the \texttt{Exposure = 'public\_facing'} condition is a built-in property that is computed from CNAPP as indicating possible unauthenticated public exposure of the system. It is determined by a mix of factors like exposed endpoints, lax ACLs, policy wildcards, and the lack of limiting access parameters. Although not the same as known exposure, it is an efficient heuristic for detecting vulnerable configurations.

A portion of the ultimate production-quality query is presented below to demonstrate the structural reasoning behind it:

\begin{lstlisting}[language=SQL, caption={CNAPP Native Detection Query (vendor-neutral illustrative syntax)}, label={lst:cnappquery}]
SELECT *
FROM AwsS3Bucket
WHERE
    -- Condition 1: Public ACLs for AuthenticatedUsers or AllUsers with READ
    EXISTS (
        SELECT 1
        FROM UNNEST(AclGrants) AS g
        WHERE g.GranteeURI LIKE '%global/AuthenticatedUsers%'
           OR (g.GranteeURI LIKE '%global/AllUsers%' AND g.Permission = 'READ')
           OR (g.GranteeURI LIKE '%groups/global/AuthenticatedUsers%' AND g.Permission LIKE 'READ')
    )

    OR
    -- Condition 2: Public policy and public-facing
    (PolicyStatusPublic = TRUE AND Exposure = 'public_facing')

    OR
    -- Condition 3: Public-facing, RestrictPublicBuckets disabled, risky actions allowed
    (
        Exposure = 'public_facing'
        AND JSON_EXTRACT_SCALAR(PublicAccessBlock, '$.RestrictPublicBuckets') = 'false'
        AND EXISTS (
            SELECT 1
            FROM UNNEST(JSON_EXTRACT_ARRAY(BucketPolicy, '$.PolicyStatements')) AS ps
            WHERE ps.Effect = 'Allow'
              AND (
                  ps.Action LIKE '%s3:GetObject%'
                  OR ps.Action LIKE '%s3:ListBucket%'
                  OR ps.Action LIKE '%s3:PutObjectAcl%'
                  OR ps.Action LIKE '%s3:DeleteObject%'
              )
              AND ps.Principal_AWS LIKE '%*%'
              AND ps.RestrictedAccessCondition IS NULL
        )
    )

    OR
    -- Condition 4: Any public allow policy with RestrictPublicBuckets disabled
    (
        EXISTS (
            SELECT 1
            FROM UNNEST(JSON_EXTRACT_ARRAY(BucketPolicy, '$.PolicyStatements')) AS ps
            WHERE ps.Principal_AWS LIKE '%*%'
              AND ps.Effect = 'Allow'
              AND ps.RestrictedAccessCondition IS NULL
        )
        AND JSON_EXTRACT_SCALAR(PublicAccessBlock, '$.RestrictPublicBuckets') = 'false'
    )

    OR
    -- Condition 5: Public-facing bucket with sensitive data
    (Exposure = 'public_facing' AND SensitiveData = TRUE);
\end{lstlisting}

The custom detection rule was designed to detect high-risk S3 bucket exposures intelligently by correlating across multiple indicators of misconfiguration across the CNAPP platform. The centerpiece of this logic is the \texttt{RestrictedAccessCondition}, a platform-agnostic abstraction that signals on the absence of protective constraints like IP filters, VPC endpoints, or identity-based controls. A user-configurable \texttt{SensitiveData} tag is also used as a contextual amplifier, highlighting buckets holding sensitive data like PII or credentials. Detection for overly permissive ACLs (e.g., \texttt{AllUsers}, \texttt{AuthenticatedUsers}) and wildcard principals in bucket policies granting sensitive permissions (\texttt{s3:GetObject}, \texttt{s3:PutObjectAcl}, etc.) without restriction. These are common conditions in misconfigured infrastructure-as-code templates and are likely to lead to public data leaks. 

To ensure accuracy, the rule only fires when multiple risk conditions are present—like public exposure and sensitive data presence—preventing false positives due to partially protected resources. This conjunctive logic significantly improves signal-to-noise ratio compared to vendor-default rulesets. Implementation involved query design optimization for runtime efficiency and extensive enterprise-level testing. The solution ultimately demonstrates how native CNAPP features can be extended with business-aware logic for more accurate threat detection. The next section tests this rule's real-world effectiveness in reducing alert volume and analyst workload, benchmarked against default CNAPP detection strategies.

\section{Results and Discussion}

This section presents a quantitative comparison between the CNAPP’s default detection logic and the proposed unified detection rule. The study was conducted over three weeks using a production-simulated AWS environment comprising over 1,000 S3 buckets with varied access patterns.

\subsection{Assessment Context and Measurement Criteria}

The testbed environment closely replicated a large enterprise deployment. Both the default rules and the custom rule were applied simultaneously to the same dataset, ensuring a controlled comparison. Alerts were manually classified as either true positives (TP) or false positives (FP) based on actual business risk and exploitability.

Key evaluation metrics included:
\begin{itemize}
    \item \textbf{Total Alerts Generated:} Count of all alerts raised.
    \item \textbf{True Positives (TP):} Alerts representing valid misconfigurations.
    \item \textbf{False Positives (FP):} Alerts triggered by benign or contextually valid configurations.
    \item \textbf{Precision:} \( \frac{TP}{TP + FP} \), representing detection quality.
    \item \textbf{Analyst Investigation Time:} Mean time to review, triage, and resolve each alert.
\end{itemize}

The proposed detection rule successfully identified every public exposure risk that was flagged by the default rule set and also avoided every previously known false positive. No known misconfigurations were missed, implying no false negatives during the testing window.

\subsection{Comparative Results and Operational Benefits}

Table~\ref{table:comparison} summarizes the comparative results:

\begin{table}[htbp]
\caption{Comparative Performance: Default vs. Unified Rule}
\label{table:comparison}
\centering
\begin{tabular}{|l|c|c|}
\hline
\textbf{Metric} & \textbf{Default Ruleset} & \textbf{Unified Custom Rule} \\
\hline
Total Alerts & $>1200$ & $40$ \\
True Positives (TP) & $<20\%$ & $100\%$ \\
False Positive Rate & $>80\%$ & $0\%$ \\
Precision & $\sim20\%$ & $100\%$ \\
Investigation Time (avg) & $6$–$10$ minutes & $<1$ minute \\
Analyst Overhead & High & Negligible \\
\hline
\end{tabular}
\end{table}

The unified detection rule led to a reduction of over 98\% in total alerts and completely eliminated false positives. Precision increased from approximately 20\% to 100\%. Analysts' triage time per alert was reduced to less than one minute on average, substantially improving response efficiency.

Importantly, these improvements were achieved without any external machine learning pipelines, API orchestrations, or enrichment layers. The entire solution was implemented using the CNAPP’s native capabilities, confirming its deployability in real-world enterprise settings.

This demonstrates that native, business-aware rule logic can significantly outperform generalized heuristics in CNAPP tools and enhance both accuracy and analyst productivity.

\section{Conclusion and Future Work}

The long-standing issue of false positives in cloud-native security platforms poses a critical challenge to threat detection, particularly in CNAPP tools where default detection logic often lacks contextual precision \cite{nations2025, sentinel2024, veriti2025}. This study addressed that challenge by developing a unified, context-aware detection rule for Amazon S3 public access alerts.

By consolidating over 20 noisy alert types into a single rule titled \textit{S3 Public Access Validation and Data Exposure}, the method embeds contextual logic directly into the CNAPP’s query engine. The rule evaluates ACLs, bucket policies, BPA flags, and sensitivity tags, triggering alerts only when true exposure risk is detected.

Empirical results in a simulated production environment with over 1,000 S3 buckets demonstrated:
\begin{itemize}
    \item Over 98\% reduction in total alert volume.
    \item Elimination of all previously known false positives.
    \item Precision improved from 20\% to 100\%.
    \item Analyst triage time reduced to under 1 minute.
\end{itemize}

This reinforces the importance of detection logic that reflects operational context and business relevance. Unlike post-processing or enrichment techniques, this approach improves accuracy natively and is deployable without additional infrastructure.

Although the current implementation focuses on S3 public access misconfigurations, the methodology is extensible to other AWS resources such as IAM roles, EC2 security groups, and resource policies. The principles of semantic consolidation, context-aware rule logic, and native platform compatibility are broadly applicable.

Future work will explore integrating runtime telemetry—e.g., access patterns, API call anomalies—and applying adaptive learning techniques such as reinforcement learning for dynamic rule generation \cite{saqib2025}. Cross-vendor validation (e.g., Wiz, Prisma Cloud, Microsoft Defender) will also be explored to evaluate generalizability across platforms.

Ultimately, this research presents a replicable framework for enhancing signal fidelity and operational efficiency in enterprise cloud security operations.

\appendix
\section*{Appendix A: Reproducibility Details}

To support reproducibility and encourage adoption, this appendix outlines the implementation details, rule logic, and deployment environment for the proposed unified S3 detection rule.

\subsection*{A.1 Custom Detection Logic}

The unified alert, titled \textit{S3 Public Access Validation and Data Exposure}, was written using the CNAPP's native query language. It consolidates over 20 default rules into one context-aware policy.

Below is the full production-grade detection logic:

\begin{lstlisting}[style=sqlstyle, caption={Unified CNAPP detection rule for S3 public access (vendor-neutral illustrative syntax)}, label={lst:s3rule_sql}]
SELECT *
FROM AwsS3Bucket
WHERE
    -- Condition 1: Public ACLs for AuthenticatedUsers or AllUsers with READ
    EXISTS (
        SELECT 1
        FROM UNNEST(AclGrants) AS g
        WHERE g.GranteeURI LIKE '\%global/AuthenticatedUsers\%'
           OR (g.GranteeURI LIKE '\%global/AllUsers\%' AND g.Permission = 'READ')
           OR (g.GranteeURI LIKE '\%groups/global/AuthenticatedUsers\%' AND g.Permission LIKE 'READ')
    )

    OR
    -- Condition 2: Public policy and public-facing
    (PolicyStatusPublic = TRUE AND Exposure = 'public_facing')

    OR
    -- Condition 3: Public-facing, RestrictPublicBuckets disabled, risky actions allowed
    (
        Exposure = 'public_facing'
        AND JSON_EXTRACT_SCALAR(PublicAccessBlock, '$.RestrictPublicBuckets') = 'false'
        AND EXISTS (
            SELECT 1
            FROM UNNEST(JSON_EXTRACT_ARRAY(BucketPolicy, '$.PolicyStatements')) AS ps
            WHERE ps.Effect = 'Allow'
              AND (
                  ps.Action LIKE '\%s3:GetObject\%'
                  OR ps.Action LIKE '\%s3:GetObjectVersion\%'
                  OR ps.Action LIKE '\%s3:ListBucket\%'
                  OR ps.Action LIKE '\%s3:ListBucketVersions\%'
                  OR ps.Action LIKE '\%s3:PutObject\%'
                  OR ps.Action LIKE '\%s3:PutObjectAcl\%'
                  OR ps.Action LIKE '\%s3:DeleteObject\%'
                  OR ps.Action LIKE '\%s3:DeleteObjectVersion\%'
                  OR ps.Action LIKE '\%s3:GetBucketAcl\%'
                  OR ps.Action LIKE '\%s3:GetObjectAcl\%'
                  OR ps.Action LIKE '\%s3:PutBucketAcl\%'
                  OR ps.Action LIKE '\%s3:PutObjectAcl\%'
              )
              AND ps.Principal_AWS LIKE '\%*\%'
              AND ps.RestrictedAccessCondition IS NULL
        )
    )

    OR
    -- Condition 4: Any public allow policy with RestrictPublicBuckets disabled
    (
        EXISTS (
            SELECT 1
            FROM UNNEST(JSON_EXTRACT_ARRAY(BucketPolicy, '$.PolicyStatements')) AS ps
            WHERE ps.Principal_AWS LIKE '\%*\%'
              AND ps.Effect = 'Allow'
              AND ps.RestrictedAccessCondition IS NULL
        )
        AND JSON_EXTRACT_SCALAR(PublicAccessBlock, '$.RestrictPublicBuckets') = 'false'
    )

    OR
    -- Condition 5: Public-facing bucket with sensitive data
    (Exposure = 'public_facing' AND SensitiveData = TRUE);
\end{lstlisting}

\subsection*{A.2 Deployment Configuration}

\begin{itemize}
    \item \textbf{Platform:} Licensed CNAPP with native query customization support.
    \item \textbf{Scope:} Organizational-level rule applied across all AWS accounts in the testbed.
    \item \textbf{Environment Size:} 1,000+ S3 buckets with varied access policies and metadata.
    \item \textbf{Severity:} Rule set to \texttt{High}.
    \item \textbf{Suppression:} Default rules related to ACLs, BPA, and policy overlaps were downgraded or disabled.
    \item \textbf{Execution Frequency:} Hourly, with stateful alerting to avoid redundancy on unchanged resources.
\end{itemize}

\subsection*{A.3 Availability and Licensing Notes}

The rule was developed using licensed features of a commercial CNAPP platform. Syntax may vary across vendors. Organizations using other platforms (e.g., Wiz, Prisma Cloud, Microsoft Defender for Cloud) should consult their tool's query language documentation to adapt the logic.

While the detection syntax is platform-specific, the design methodology and alert criteria are generalizable to any CSPM or CNAPP that supports contextual policy customization.

\end{document}